\documentclass[epj]{svjour}
\usepackage{latexsym}
\usepackage{graphics}
\usepackage{epsf}
\begin{document}
\title{Scattering theory and ground-state energy of
Dirac fermions in graphene with two Coulomb impurities}
\author{Denis Kl\"opfer\inst{1} \and Alessandro De Martino\inst{2}
\and Davron Matrasulov\inst{3} \and Reinhold Egger\inst{1} }                   
\institute{Institut f\"ur Theoretische Physik, Heinrich-Heine-Universit\"at,
D-40225 D\"usseldorf, Germany \and 
Department of Mathematics, City University London,
London EC1V 0HB, United Kingdom \and 
Turin Polytechnic University in Tashkent, 17 Niyazov Street,
100095 Tashkent, Uzbekistan }
%
%
\abstract{
We study the physics of Dirac fermions in a gapped 
graphene monolayer containing two Coulomb impurities.  For the case of equal
impurity charges, we discuss the ground-state energy
using the linear combination of atomic orbitals (LCAO) approach.  
For opposite charges of the Coulomb centers, an electric dipole 
potential results at large distances. We provide a nonperturbative 
analysis of the corresponding low-energy scattering problem. 
 }
\authorrunning{D. Kl\"opfer \textit{et al.}}
\titlerunning{Dirac fermions with two Coulomb impurities}
\maketitle
\section{Introduction}
\label{intro}

The amazing properties of graphene monolayers have attracted
much attention in physics and material science over the past decade.
It is by now well established that in the vicinity of the charge 
neutrality point, electronic quasi-particle excitations correspond to 
two-dimensional (2D) Dirac fermions \cite{rmp1}.  
In the presence of a bulk gap, $\Delta$, these are massive fermions.
The gap can be caused by various different mechanisms.  
To list just a few, let us mention strain-engineered gaps \cite{strain}, 
artificially designed spin-orbit interaction processes \cite{soi}, and strong 
Coulomb effects causing an excitonic insulator phase \cite{rmp2}.
Additional ways to open a gap come from superlattices that arise when
the graphene layer is deposited on a substrate 
\cite{pono,levitov2}, or simply due to the transverse finite-size gap in 
graphene nanoribbons \cite{rmp1}.  Graphene monoloyers thus imply new 
possibilities for experimentally accessing relativistic quantum phenomena 
within a readily available table-top setting.

A prime example for such a relativistic quantum effect concerns
\textit{supercriticality}: In the presence of a Coulomb impurity of charge
$Q=Ze$ 
 (where the electron has charge $-e$), 
the attractive $1/r$ potential induces bound states with energy $E$ 
inside the gap, $|E|<\Delta$.
For sufficiently large $Z>Z_c$, these bound states are predicted to
``dive" into the filled Dirac sea  
\cite{rmp2,khalilov,novikov,pereira,levitov,gamayun,denis,exp1}, 
whereby the nucleus captures an electron to reduce its charge.
In conventional realizations, the large value of the critical
charge, $Z_c\approx 170$, renders the experimental observation of 
supercriticality prohibitively difficult
 \cite{greiner,popov}.  In graphene, the Fermi velocity $v_F\ll c$ 
takes the role of the speed of light $c$, and therefore the 
effective fine structure constant becomes of order $\alpha=e^2/\hbar
v_F\approx 1$ instead of $e^2/\hbar c\simeq 1/137$. 
As a result, in graphene one expects a much smaller value for the
critical charge, $Z_c\approx 1$.  Indeed, this
reduction in $Z_c$ has already allowed one to 
observe supercriticality by tunneling spectroscopy
measurements in graphene monolayers \cite{exp2,exp3}. In those
experiments, a Coulomb center was artificially created 
by pushing together charged Co adatoms \cite{exp1,exp2} with the help
of a scanning tunneling microscope (STM) tip.  A similar procedure has
also been successfully used for Ca adatoms \cite{exp3}. 
Importantly, by local gating it is possible to vary the charge $Q$ 
of the resulting cluster (including the sign) in a controlled manner.

These recent developments allow one to envision new types of 
artificial molecules, composed of $N_Z$ nuclei with designed
charges and $N$ electrons bound to them.  Since the
electron dynamics is now governed by the Dirac equation, such
molecules may realize a  relativistic 2D counterpart to conventional
molecules, with potentially much stronger correlation effects.  
Note that for different signs of the nuclear charges, 
one has a polar molecule.  We here study the 
simplest problem of this class, which is defined by just one 2D
Dirac fermion, $N=1$, in the field of two static Coulomb centers, $N_Z=2$, 
separated by a distance $R$.  The Coulomb centers have 
charges $Q_1$ and $Q_2$, where we restrict ourselves to 
symmetric and antisymmetric configurations, $Q_2=\pm Q_1$.
The symmetric case represents a relativistic 2D cousin of 
the well-known $H_2^+$ problem \cite{ctan}, 
while the antisymmetric case corresponds to a polar molecule, i.e.,
an electric dipole.  The main theoretical difficulty in treating 
this problem is that the 2D Dirac equation with a two-center potential 
does not separate in any known orthogonal coordinate system.

First, for equal charges, $Q_{1,2}=Q=Ze$, we have a symmetric
two-center problem for Dirac fermions in graphene.  As recently
discussed in Ref.~\cite{gusynin}, one then encounters
a supercritical instability again. Indeed, for $R\to 0$, the problem
reduces to a single Coulomb impurity with charge $2Q$, 
while for $R\to \infty$, we recover the
charge-$Q$ single-impurity problem.  Assuming a subcritical value of $Q$ 
such that $1/2<\zeta<1$, with $\zeta=Z/Z_c$, one expects that by 
pushing together the Coulomb centers, supercriticality emerges below 
some critical distance, $R<R_{\rm cr}$. The idea to induce supercriticality
by bringing together two Coulomb centers has already been discussed in the 
1970s for colliding heavy nuclei \cite{gerstein,rafelski,zeldovich},
and is implicitly behind the recent graphene experiments  \cite{exp2,exp3}.
The critical distance for the graphene case was
computed in Ref.~\cite{gusynin} by an asymptotic matching procedure,
leading to the transcendental equation 
\begin{equation}\label{guseq}
2\sqrt{1-\zeta^2}-1= \frac{zK'_{i\gamma}(z)}{K_{i\gamma}(z)} , \quad z=
2\sqrt{\zeta R_{\rm cr}/R_\Delta},
\end{equation}
with $\gamma=\sqrt{4\zeta^2-1}$, $R_\Delta=\hbar v_F/\Delta$, and the 
Macdonald  function $K_{i\gamma}(z)$ with imaginary order 
\cite{gradst,abramowitz}.
We here develop an alternative description of this phenomenon based on the 
``linear combination of atomic orbitals" (LCAO) approach, which is
commonly used in molecular physics \cite{ctan}.  The 
corresponding two-center Dirac problem in three dimensions (3D) has
been studied by the LCAO technique in Refs.~\cite{davron2,bondarchuk}.  

For opposite charges,  $Q_2=-Q_1=Q$, the two-center potential reduces for large
distances, $r\gg R$, to the potential of an \textit{electric dipole}  
with dipole moment $P=QR$,
\begin{equation} \label{pointdipole}
V_d(r,\theta)=-\frac{eP\cos\theta}{r^2},
\end{equation}
where $\theta$ is the azimuthal angle between ${\bf r}$ and the dipole axis.
$V_d$ is referred to below as the "point-like dipole" potential.
A similar $1/r^2$ potential also describes 
the conical singularities near graphene wrinkles \cite{wrinkle}. 
Recently, we have presented a related study of the dipolar 
two-center problem in
graphene \cite{dipolPRL}, where we have analyzed the bound states induced
by the dipole, see also Sec.~\ref{sec4a} below.
It turns out that an arbitrarily weak dipole can already bind infinitely many 
states.  This is in contrast to 3D Schr\"odinger fermions, where the
dipole is able to capture bound states 
only when $P$ exceeds a finite critical strength
 \cite{AK,davron,dipol0,dipol1,num-efimov}.  However, for the 2D
Schr\"odinger case, the critical dipole strength vanishes as well \cite{dipol1}.
In Ref.~\cite{dipolPRL}, we have -- albeit very briefly --  also discussed the
scattering problem in graphene, using the dipolar two-center Dirac equation  
within the perturbative Born approximation.  
The resulting scattering state implies 
a completely isotropic transport cross-section.  Within the 
Born approximation, this predicts that charge transport is
independent of the angle between current flow and dipole direction.
In the present work, we elaborate on the scattering problem also
 beyond the Born approximation, including the 
nonperturbative low-energy regime.

The structure of the remainder of this article is as follows.
In Sec.~\ref{sec2}, we introduce the model and discuss its symmetries.
Sec.~\ref{sec3} provides a discussion of our Dirac-LCAO calculations
for the symmetric two-center problem.
In Sec.~\ref{sec4}, we then describe the scattering theory results
for the dipolar potential with opposite nuclear charges.
We finally offer some concluding remarks in Sec.~\ref{conc}.

\section{Model}\label{sec2}

Throughout this paper, we study 2D Dirac fermions 
 with a mass gap $\Delta$ in the presence of a static
two-center potential.  Using units with $\hbar=e=1$, the Hamiltonian reads
\cite{rmp1}
\begin{equation}\label{eq1}
H= -iv_F\partial_x \sigma_x-iv_F\partial_y\sigma_y +\Delta \sigma_z + V.
\end{equation}
The Pauli matrices $\sigma_{x,y,z}$ act in sublattice space,
corresponding to the two atoms in the elementary cell of graphene's
honeycomb lattice. Following standard arguments, we consider a single $K$
point (``valley") and a single electron spin projection only \cite{rmp1}.
It is worth mentioning that Eq.~(\ref{eq1}) also describes
other Dirac materials, e.g., the ``molecular graphene" resulting from 
the deposition of CO molecules on a copper surface as described 
in Ref.~\cite{gomes},
or the recently discovered surface states of topological insulator materials
such as Bi$_2$Se$_3$ or Bi$_2$Te$_3$ \cite{hasan}.
The Hamiltonian (\ref{eq1}) acts on states with 
two spinor components, $\Psi({\bf r})=(\eta,\chi)^T$.
With the nuclear charges $Z_{1,2}$,
the two-center potential $V$ in Eq.~(\ref{eq1}) reads 
\begin{equation}\label{twocenter}
V(x,y)= -\frac{Z_1}{r_1}-\frac{Z_2}{r_2},
\end{equation}
with the distances $r_{1,2}$ of the electron to the two Coulomb centers.
Assuming that the centers are at $y=0$ and $x=\mp R/2$, resp., we have
\begin{equation}\label{r12def}
r_{1,2} = \sqrt{(x\pm R/2)^2+y^2}.
\end{equation}
The dipole case is realized for $Z_2=-Z_1=Z$, with dipole moment $P=ZR$.
For $Z_1=Z_2=Z$, we instead have a symmetric two-center problem.

Let us next discuss the symmetry properties of this system. 
For $Z_1=Z_2$, the Hamiltonian is 
invariant under a $\pi$-rotation exchanging the two Coulomb centers. 
Indeed, with the total angular momentum operator, 
\begin{equation}\label{jzdef}
J_z =-i\partial_\theta +\frac{\sigma_z}{2}, 
\end{equation}
the unitary operator ${\cal R}_\pi=e^{i\pi J_z}$ generates
the shift $\theta\to \theta+\pi$ and commutes with the Hamiltonian, 
$[H,{\cal R}_\pi]_-=0$.  The spinor is thereby mapped to 
\begin{equation}\label{symmetricmap}
\Psi(x,y)\to {\cal R}_\pi \Psi= i \sigma_z \Psi(-x,-y).
\end{equation}
We note that the single-valley Hamiltonian (\ref{eq1}) is 
\textit{not}\ invariant under the reflection $x\to -x$, 
which maps left- to right-handed quasi-particles.

In the dipolar case, $Z_2=-Z_1$, the Hamiltonian does not
have the above symmetry, but instead it maps to $-H$ by the
unitary transformation $U=\sigma_x {\cal R}_x$, where ${\cal R}_x$ 
performs the reflection $x\to -x$.  Indeed, we find $UHU^\dagger= -H$, 
which implies a particle-hole-like symmetry of the entire spectrum.  
By virtue of this unitary transformation, out of an 
eigenstate $\Psi_E(x,y)$ at energy $E$, one immediately obtains a
partner state at energy $-E$,
\begin{equation}\label{unitary}
\Psi_{-E}(x,y) = U \Psi_E(x,y)= \sigma_x \Psi_E(-x,y).
\end{equation}
All solutions to the dipolar two-center Dirac equation
therefore come in $\pm E$ pairs, and
it is sufficient to study only, say, $E<0$. 
We note that there are no $E=0$ solutions \cite{dipolPRL}.

\section{LCAO approach for  symmetric two-center problem}
\label{sec3}

In this section, we present ground-state results 
obtained from the LCAO approach for the symmetric
two-center Dirac problem in 2D. The corresponding potential $V$ is
given in Eq.~(\ref{twocenter}) with $Z_1=Z_2=Z$.
The LCAO approximation solves the Dirac equation in a truncated
subspace, where only the lowest single-impurity bound state
near each center is retained.  This approximation
is expected to yield accurate ground-state energies for large $R$ 
\cite{davron2,bondarchuk}, where the molecular ground state
is well approximated in terms of atomic orbitals. 
In addition, as we show below, 
the exact result for $R\to 0$ is also captured 
by the LCAO solution.

\subsection{Single-impurity ground state}

Since the LCAO method employs a superposition of states localized near one of
the centers, we first summarize the known single-impurity 
solution for the lowest bound state \cite{khalilov,novikov,gamayun}.  
Taking a single impurity of charge $Z$, 
i.e., using $V=-Z/r$ in Eq.~(\ref{eq1}), 
the lowest bound state has the energy $\xi\Delta$, with
\begin{equation}\label{eps0}
\xi =  \sqrt{1- 4Z^2}. 
\end{equation}
In the absence of short-distance regularization, the supercritical 
threshold is  reached at $Z_c=1/2$ \cite{gamayun}, and 
we assume $Z<Z_c$ henceforth. 
The corresponding spinor is an eigenstate of the total angular 
momentum operator $J_z$ in Eq.~(\ref{jzdef}), with
eigenvalue $1/2$. 
Using the lengthscale $R_\Delta=\hbar v_F/\Delta$, and $\xi$ 
in Eq.~(\ref{eps0}), it reads
\begin{eqnarray}\label{singleimp}
\Psi_0 (r,\theta) & =& \frac{2Z}{\sqrt{\pi \Gamma(1+\xi)} R_\Delta} 
\left(\frac{4Zr}{R_\Delta}\right)^{(\xi-1)/2} \\
\nonumber &\times& 
e^{-2Zr/R_\Delta} \left( \begin{array}{c} \sqrt{1+\xi} \\ ie^{i\theta}  
\sqrt{1-\xi} \end{array} \right),
\end{eqnarray}
where $\Gamma(z)$ is the Gamma function. 

\subsection{LCAO scheme}

Using the kinetic (Dirac) Hamiltonian $H_0=H-V$ in Eq.~(\ref{eq1}), 
we first re-write the Hamiltonian as
\begin{equation}\label{qeffintro}
H= H_0 - Z_{\rm eff}\left(\frac{1}{r_1}+\frac{1}{r_2}\right)
-\frac{\delta Z_1}{r_1} -\frac{\delta Z_2}{r_2},
\end{equation}
where $\delta Z_{1,2}=Z_{1,2}-Z_{\rm eff}$, and $r_{1,2}$
has been defined in Eq.~(\ref{r12def}). (We will put $Z_1=Z_2$ later on.)
While Eq.~(\ref{qeffintro}) is, of course, exact for arbitrary effective 
charge $Z_{\rm eff}$, 
the LCAO approximation obtains a ground-state 
energy, $E$, that still depends on the value
of $Z_{\rm eff}$.  The final LCAO ground-state energy
is then obtained by minimizing
$E(Z_{\rm eff})$ with respect to the variational parameter $Z_{\rm eff}$.

Following the standard LCAO approach \cite{ctan}, we expand the 
ground state $|\Phi\rangle$
of Eq.~(\ref{qeffintro}) in terms of atomic orbitals, 
$|1\rangle$ and $|2\rangle$,
centered near the Coulomb impurity at $(\mp R/2,0)$, 
respectively, i.e., $|\Phi\rangle = v_1 |1\rangle+ v_2|2\rangle.$
The atomic orbitals are chosen as single-impurity states (\ref{singleimp}), 
where the energy $\xi\Delta$ follows from Eq.~(\ref{eps0})
with $Z\to Z_{\rm eff}$.
The Dirac equation is thereby reduced to a linear $2\times 2$ equation 
for $(v_1,v_2)$, and the energy $E=E(Z_{\rm eff})$ follows from the 
condition
\begin{equation}\label{detc}
{\rm det} \left(\begin{array}{cc}  H_{11}-E & H_{12}-SE \\
H_{21}-SE & H_{22}-E \end{array}\right) = 0,
\end{equation}
with the overlap integral
\begin{equation}\label{overlap}
S= \langle 1|2\rangle=\langle 2|1\rangle.
\end{equation}
Note that the single-impurity state (\ref{singleimp}) is normalized, 
and therefore we have $\langle 1|1\rangle= \langle 2|2\rangle=1.$ 
Defining the Coulomb integral,
\begin{equation}\label{coulint}
C = \langle 1|r_2^{-1} |1 \rangle = \langle 2 |r^{-1}_1|2\rangle,
\end{equation}
and the resonance integral,
\begin{equation}\label{resint}
A=\langle 1|r^{-1}_{1,2}|2\rangle=
 \langle 2|r_{1,2}^{-1}|1\rangle,
\end{equation}
and using the relation 
\begin{equation}
\langle 1|r^{-1}_{1}|1\rangle=4 Z_{\rm eff} \Delta/\xi,
\end{equation}
all matrix elements in Eq.~(\ref{detc}) can be written in compact form,
\begin{eqnarray} 
H_{11}&=& \xi\Delta - 4Z_{\rm eff}\delta Z_1\Delta/\xi - Z_2 C, \\  \nonumber
H_{12}&=& \xi S\Delta -(\delta Z_1  +Z_2 ) A, \\ \nonumber
H_{21}&=& \xi S\Delta- (Z_1 +\delta Z_2 ) A,\\ \nonumber
H_{22}&=& \xi\Delta -  Z_1 C- 4 Z_{\rm eff}\delta Z_2\Delta/\xi.
\end{eqnarray}
While in the 3D Dirac problem, the quantities $S$, $C$, and $A$ can be
directly evaluated \cite{davron2,bondarchuk},
the 2D case is, unfortunately, more involved.

\subsection{Overlap, Coulomb, and resonance integrals}

In order to compute the quantities $S$, $C$, and $A$, it is useful
to employ elliptic coordinates.  Denoting the distances of the electron
from the two centers by $r_{1,2}$, see Eq.~(\ref{r12def}), 
elliptic coordinates are defined by \cite{gradst} 
\begin{equation}
\mu=\frac{r_1+r_2}{R}\in [1,\infty),\quad \nu=\frac{r_1-r_2}{R}\in [-1,1],
\end{equation}
where the standard cartesian coordinates are
\begin{equation}
x=\frac{R}{2}\mu\nu,\quad y=\pm \frac{R}{2}\sqrt{(\mu^2-1)(1-\nu^2)}.
\end{equation}   
The $\pm$ sign is chosen according to whether ${\bf r}=(x,y)$ is in the
upper or lower half-plane: the transformation between cartesian
and elliptic coordinates is only one-to-one in each half-plane.   
The segment $\mu=1$ with $-1\le \nu\le 1$
then corresponds to the points on the $x$-axis between $-R/2$ and $+R/2$,
while the regions $x\le -R/2$ ($x\ge R/2$) are covered by $\nu=-1$ ($\nu=1$)
with $1\le \mu<\infty$, respectively.  
We note that the Jacobian determinant,
\begin{equation}
{\rm det} J= \mp \frac{R^2}{4} \frac{\mu^2-\nu^2}{\sqrt{(\mu^2-1)(1-\nu^2)}},
\end{equation} 
is singular along the full $x$-axis.
In terms of elliptic coordinates, the sought quantitites ($S$, $C$, and $A$) 
are thereby expressed as integrals over $\mu$ and $\nu$. 

Let us start with the overlap integral $S$ in Eq.~(\ref{overlap}).
Using the abbreviations $\xi=\sqrt{1-4Z_{\rm eff}^2}$,
\begin{equation}
{\cal N} = \frac{4u^{1+\xi}}{\pi \Gamma(1+\xi)}, \quad
u = \frac{2R Z_{\rm eff}}{R_\Delta},
\end{equation}
it takes the form 
\begin{eqnarray}\nonumber
S &=& {\cal N} \int_1^\infty d\mu \int_{-1}^1 d\nu 
\frac{(\mu^2-\nu^2)^{(\xi-1)/2} e^{-u\mu}}{\sqrt{(\mu^2-1)(1-\nu^2)}}
\\ \label{sinit} &\times& \left[\mu^2-1+\xi(1-\nu^2)\right].
\end{eqnarray}
By virtue of the auxiliary relation
\begin{eqnarray}
&& \int_{-1}^1 d\nu (1-\nu^2)^{\pm 1/2} \left(1-\frac{\nu^2}{\mu^2}\right)^s 
\\ \nonumber && = \frac{2\pi}{3\pm 1} F\left( \frac12,-s;\frac{3\pm 1}{2};
\frac{1}{\mu^2}\right),
\end{eqnarray}
the $\nu$-integration can be performed.   
In the next step, we employ a standard series representation for 
the hypergeometric function $F(a,b;c;d)$
\cite{gradst}, where the resulting summation commutes
with the $\mu$-integration in Eq.~(\ref{sinit}).  After this
integration, we encounter the function
\begin{equation}\label{aux2}
{\cal I}(s,u)= \int_1^\infty d\mu \frac{\mu^s e^{-u \mu}}{\sqrt{\mu^2-1}},
\end{equation}
which can be evaluated in closed form. 
With the Pochhammer symbol $(a)_n$, recursively
defined by $(a)_n/(a)_{n-1}=a+n-1$ and $(a)_0=1$ \cite{gradst}, 
we arrive at a rapidly convergent series, 
\begin{eqnarray}\nonumber
S&=&\pi {\cal N} \sum_{n=0}^\infty \frac{(1/2)_n}{n!} \Biggl \{
\frac{ \left( -\frac{1+\xi}{2} \right)_{n}  }{(1)_n} 
{\cal I}(\xi+1-2n,u) \\ &-& \frac{1-\xi}{2}  \label{finals}
\frac{ 
\left( \frac{1-\xi}{2} \right)_{n}  }{(2)_n} {\cal I}(\xi-1-2n,u) \Biggr\}. 
\end{eqnarray}
By very similar steps, we also obtain the Coulomb integral,
\begin{eqnarray}\nonumber
C&=& \frac{2\pi{\cal N}}{R}  \sum_{n=0}^\infty \frac{(1/2)_n}{n!} \Biggl 
\{ \frac{ \left( \frac{-\xi+1}{2} \right)_{n}  }{(1)_n} {\cal I}(\xi-2n,u) 
\\  && - \frac{1-\xi}{2} \label{finalc}
\frac{ \left( \frac{3-\xi}{2} \right)_{n}  }{(2)_n} {\cal I}(\xi-2-2n,u) 
\Biggr\}. 
\end{eqnarray}
Concerning the resonance integral $A$, one has to proceed in a 
different manner. Elliptic coordinates yield the expression
\begin{equation}
A = \frac{ 4 {\cal N}}{R} \int_1^\infty d\mu 
\int_{-1}^1 d\nu \frac{(\mu+\nu)^{\xi} e^{-u(\mu+\nu)} } {
\sqrt{(\mu^2-1)(1-\nu^2)} }.
\end{equation}
Expanding $(1+\nu/\mu)^\xi= \sum_{n=0}^\infty \left( 
\begin{array}{c} \xi  \\ n \end{array}\right) (\nu/\mu)^n$, 
the $\nu$-integrals are done using
\begin{equation}
\int_{-1}^1 d\nu \frac{e^{-u\nu} \nu^n}{\sqrt{1-\nu^2}} 
= (-1)^n \pi 
\frac{\partial^n I_0(u)}{\partial u^n} ,
\end{equation}
where $I_0$ is the modified Bessel function \cite{gradst}. 
The subsequent $\mu$-integration then leads to expressions as
in Eq.~(\ref{aux2}), and we get the series representation
\begin{equation}
A = \frac{4\pi {\cal N}}{R} \sum_{n=0}^\infty (-1)^n \left(
\begin{array}{c}\xi \\ n \end{array}\right)  
{\cal I}(\xi-n,u) \frac{\partial^n I_0(u)}{\partial u^n} ,
\end{equation}
which is also rapidly convergent.  We now put $Z_1=Z_2=Z$ 
and turn to the LCAO results for the ground-state energy.

\subsection{LCAO results}

\begin{figure}
\begin{center}
\vspace*{0.5cm}
\epsfxsize=0.96\columnwidth
\epsffile{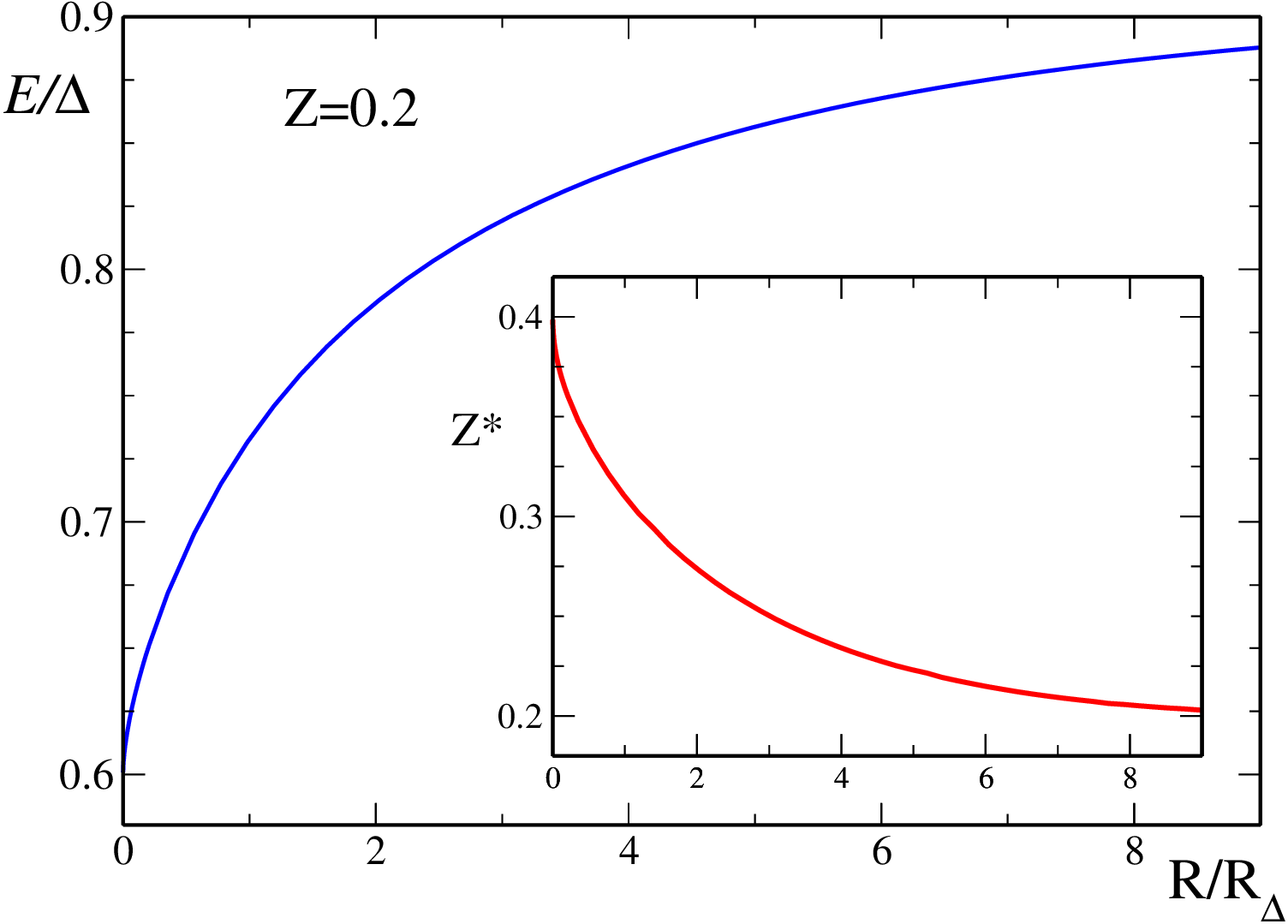}
\caption[]{Main panel: 
LCAO result for the ground-state energy, $E$, vs 
impurity distance $R$ (which is given in units of
$R_\Delta=\hbar v_F/\Delta$), for the two-center potential
with $Z_1=Z_2=Z=0.2$. 
Inset: Optimal choice for the variational parameter, $Z_{\rm eff}=Z^*$, 
determining the LCAO ground state, as a function of $R$. 
\label{fig1}}
\end{center}     
\end{figure}

Using the above series representations for $S$, $C$, and $A$, it is 
numerically straightforward to obtain the LCAO estimate for the 
ground-state energy $E(Z_{\rm eff})$ for given $Z_{\rm eff}$.
We then determine the minimal energy, realized for $Z_{\rm eff}=Z^*$, 
where the numerical search is aided by noting that 
$E(Z_{\rm eff})$ depends quadratically on $Z_{\rm eff}-Z^*$. 
The optimal value, $Z^*$, is shown in the
inset of Fig.~\ref{fig1}.  The main panel of Fig.~\ref{fig1} gives
the corresponding ground-state energy for $Z=0.2$, where supercriticality
is never reached since we have chosen a value with $2Z<Z_c=1/2$, i.e.,
$E(R)>-\Delta$ for all values of the impurity distance $R$.  
We also observe that the LCAO ground-state energy, $E(R)$, matches the 
expected single-impurity values in Eq.~(\ref{eps0}) in both limits, namely
(i) for $R\to \infty$ with impurity charge $Z$, where we have
two decoupled copies of the single-impurity problem,
and (ii) for $R\to 0$, where
both centers conspire to form a single Coulomb impurity of charge $2Z$.
Furthermore, the inset illustrates that the 
optimal effective charge $Z^*$ nicely matches both limits as well.

\begin{figure}
\begin{center}
\vspace*{0.5cm}
\epsfxsize=0.96\columnwidth
\epsffile{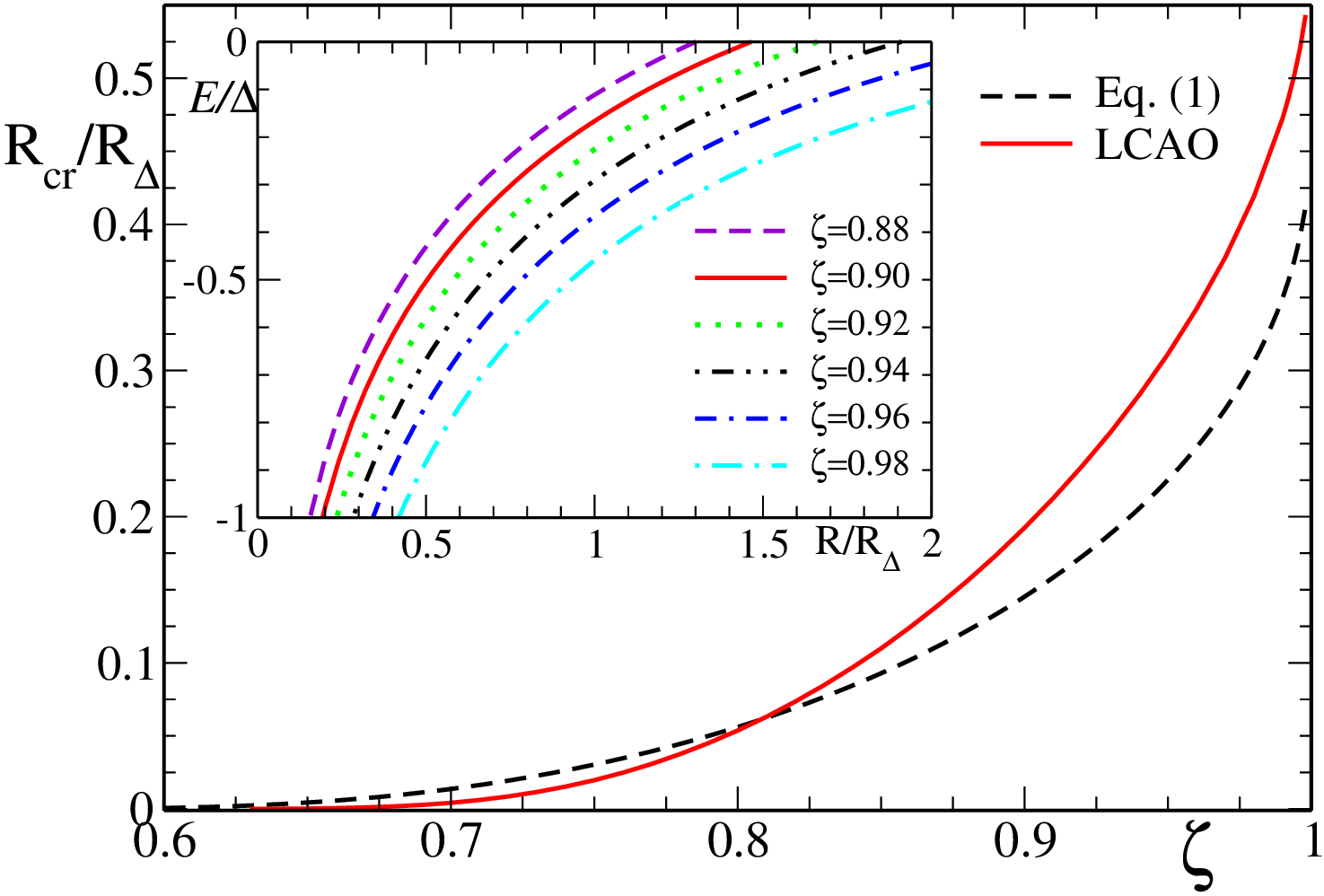}
\caption[]{ Main panel: Critical distance $R_{\rm cr}$  
vs $\zeta=Z/Z_c$, for the symmetric two-center problem.
The red curve gives the LCAO result, and the black curve is
the corresponding asymptotic matching result (\ref{guseq}).
Inset: LCAO ground state energy vs $R$, for various $\zeta$. 
Once $E(R)$ reaches $-\Delta$, the level ``dives'' into the lower
continuum and turns supercritical.
\label{fig2}}
\end{center}     
\end{figure}

Choosing larger $Z$ such that $\zeta=Z/Z_c=2Z$ is within the 
bounds $1/2<\zeta<1$, the supercritical regime can be realized by decreasing $R$ through a
transition value, $R=R_{\rm cr}$.  At the critical distance, 
the ground-state energy reaches the Dirac sea, 
$E(R_{\rm cr})=-\Delta,$
and for $R<R_{\rm cr}$, the two-center system with subcritical 
individual impurity charge becomes supercritical. 
The LCAO prediction for the critical distance $R_{\rm cr}$ is shown
as a function of $\zeta$ in the main panel of Fig.~\ref{fig2}, where 
the inset illustrates our strategy for obtaining $R_{\rm cr}$.

The LCAO results in Fig.~\ref{fig2} are rather similar to the 
predictions of Eq.~(\ref{guseq}) and indicate that, in practice,
$Z$ has to be chosen quite close to $Z_c=1/2$,
since otherwise $R_{\rm cr}$ becomes extremely small.   This conclusion
seems also in agreement with the reported experimental observations
of supercriticality \cite{exp2,exp3}, where different ions first had to 
be pushed closely together, thereby forming charged clusters, 
before supercriticality appears.

\section{Dipolar two-center potential} \label{sec4}

In this section, we turn to the dipolar case, $Z_2=-Z_1=Z$ 
in Eq.~(\ref{twocenter}). 
We first analyze the conditions for bound states in this potential,
thereby also summarizing those results of Ref.~\cite{dipolPRL} that are
relevant for the subsequent discussion.  
We then turn to the scattering problem.  After presenting the
general scattering state for the Dirac problem in Sec.~\ref{sec4b},
we study the results of the Born approximation
in Sec.~\ref{sec4c}. This approximation holds when the energy 
of the scattering state does not approach the band edge; otherwise
 a nonperturbative analysis
is required and will be given  in Sec.~\ref{sec4d}.

\subsection{Bound state spectrum} \label{sec4a}

We start by analyzing the possibility of bound states induced by 
the dipolar two-center potential.  For the corresponding Schr\"odinger
case, a scaling argument shows \cite{dipol1} that all energy eigenvalues
must be of the form $E_{\rm Schr}=(mR^2)^{-1} {\cal F}(mP)$
[recall that we use units with $\hbar=e=1$],
where $m$ is the mass of the electron, the dipole moment
is $P=ZR$, and ${\cal F}$ denotes a dimensionless
one-parameter scaling function.
The critical dipole strength allowing for bound states follows
from the condition ${\cal F}=0$, which does not involve $Z$ and $R$ 
separately but only the combination $P=ZR$.  This implies that
both the two-center potential and the point-like dipole form (with $P=ZR$)
lead to the same critical coupling. For the 2D case, this critical
value is zero \cite{dipol1}.

Repeating this scaling argument for the Dirac case,
we see that the energy eigenvalues can be written as
$E_{\rm Dirac} = \Delta {\cal G}(P\Delta/v_F^2, R\Delta/v_F),$
with a \textit{two-parameter} scaling function ${\cal G}$.
The condition for bound states now  becomes ${\cal G}=\pm 1$, which implies
that, in general, the critical dipole coupling still depends on the
impurity distance $R$.  However,  for $R\Delta/v_F\gg 1$, 
${\cal G}$ effectively becomes a one-parameter scaling function again,
and the critical coupling is independent of $R$. In this subsection,
we assume that this limit is realized.

Noting that the entire spectrum is particle-hole symmetric,
our analysis in Ref.~\cite{dipolPRL} showed
that weakly bound states of energy $E=\pm(\Delta-\epsilon)$,
for $0<\epsilon\ll \Delta$, 
are arranged in infinite ``towers''.  In each tower, all bound states
have the same ``angular momentum'', $j=0,1,2,\ldots$; the latter
differs from true angular momentum which is not
conserved due to the lack of isotropy.
Different towers are also labeled by a
parity eigenvalue, $\kappa=\pm$, describing the symmetry of the 
solution under the spatial inversion $x\to -x$, where the dipole 
points along the $x$-axis. Allowed towers have 
to satisfy $j+\kappa\ge 0$.
For given $(j,\kappa)$, the respective tower of bound states 
only exists if $P$ exceeds a critical value, $P>P_{j,\kappa}$.
Once this is the case, the dipole hosts infinitely many
states belonging to this tower.  Remarkably, since
$P_{0,+}=0$, one always has at least one tower. 
 For a mathematically rigorous discussion of these points, 
we refer to Ref.~\cite{cuenin}.  The threshold couplings $P_{j,\kappa}$
are ordered as $P_{0,+}<P_{1,-}<P_{1,+}<P_{2,-}<\cdots$,
and using the approach of Ref.~\cite{AK}, 
we found that for $j>0$ a very good approximation is given by
\cite{dipolPRL}
\begin{equation}\label{pjk}
P_{j,\kappa} \simeq \frac{\Gamma^4(1/4)}{64\pi} \frac{v^2_F}{\Delta} \left[
\left(2j+\frac{\kappa}{2}\right)^2-\frac{1}{6\pi} \right].
\end{equation}
 Bound states within the same tower 
obey the  scaling hierarchy ($n=1,2,\ldots$)
\begin{equation}\label{efimov}
\frac{\epsilon_{n+1}}{\epsilon_{n}}= \exp \left(-\frac{2\pi}{
s_{j,\kappa}} \right) , 
\end{equation}
with the numbers ($P>P_{j,\kappa}$)   
\begin{equation}\label{ssdef}
s_{j,\kappa} \simeq \left\{ \begin{array}{ll} 
\sqrt{2} P\Delta, & (j,\kappa)=(0,+),\\
 0.956 \sqrt{(P-P_{j,\kappa})\Delta},& j>0. \end{array}\right.
\end{equation}
For $n\to \infty$, all bound states accumulate near the gap edges 
according to the universal Efimov scaling law (\ref{efimov}).
This law also describes bound states of three bosons in free space
\cite{efimov,efimov2,gogolin}. 

\subsection{Scattering state}\label{sec4b}

Let us now turn to scattering states.
For an incoming plane wave with momentum ${\bf k}=k(\cos\phi_{\bf k},
\sin\phi_{\bf k})$, using $\sigma=\pm$ 
to distinguish positive and negative energy solutions, 
the Dirac scattering state has the energy 
\begin{equation}
E_{{\bf k},\sigma}= \sigma \sqrt{\Delta^2+v_F^2k^2}.
\end{equation}
The asymptotic form of the state for $r\gg R$ 
contains an outgoing spherical wave,
\begin{eqnarray} \label{scatt}
 \Psi_{{\bf k},\sigma}(r,\theta) &\simeq &
  e^{ i{\bf k}\cdot {\bf r}} U_{{\bf k},\sigma}
+ f(\theta,\phi_{\bf k}) \frac{e^{ikr}}{\sqrt{-ir}} U_{k\hat e_r,\sigma}, 
\end{eqnarray}
with ${\bf r}=r \hat e_r$,  $\hat e_r = (\cos\theta,\sin\theta)$, and
the spinor \cite{rmp1}
\begin{equation}
U_{{\bf k},\sigma}  = \frac{1}{ \sqrt{ 2|E_{{\bf k},\sigma}| } }
\left(\begin{array}{c} \sqrt{|E_{{\bf k},\sigma}+\Delta|}
e^{-i\phi_{\bf k}/2}\\ \sigma\sqrt{|E_{{\bf k},\sigma}-\Delta|}
e^{i\phi_{\bf k}/2} \end{array} \right).
\end{equation}
Below, we first discuss the Born approximation for the
scattering amplitude $f(\theta,\phi_{\bf k})$, followed by
a nonperturbative treatment. This becomes necessary when the energy 
approaches the gap, but is analytically possible
only for the point-like dipole potential, $V_d$.  
For such a $1/r^2$ potential, it is well-known that 
a short-distance regularization scheme is required
to prevent the usual fall-to-the-center problem \cite{popov2}. 

\subsection{Born approximation}\label{sec4c}

Treating the scattering problem within perturbation theory, 
the outgoing part of the scattering state reads 
\cite{novikov,zazu} 
\begin{eqnarray}\label{outstat}
\Psi_{{\bf k},\sigma}^{\rm (out)} ({\bf r}) & = &
-\int dx' dy' G_{{\bf k},\sigma}({\bf r}-{\bf r}') \Bigl(
-i\sigma_x\partial_{x'}\\ \nonumber &-&i\sigma_y\partial_{y'}  
+\Delta \sigma_z
+ E_{{\bf k},\sigma} \Bigr) V({\bf r'}) e^{i{\bf k}\cdot {\bf r}'}
U_{{\bf k}, \sigma},
\end{eqnarray}
with ${\bf r}'=(x',y')$ and 
\begin{equation}
G_{{\bf k},\sigma}({\bf r}) = \frac{i\sigma}{4} H_0^{(1)}(kr)
\simeq \frac{i\sigma}{2} \frac{e^{ikr-i\pi/4}}{\sqrt{2\pi kr}},
\end{equation}
where the second expression uses the asymptotic form of the Hankel
function $H^{(1)}_0$ \cite{gradst}.  Here, and in the remainder of the
paper, we often use units with $v_F=1$.
The results in this subsection are obtained by using the full
two-center potential $V({\bf r})$ with $Z_2=-Z_1$, see Eq.~(\ref{twocenter}).

The Born approximation for 
the scattering amplitude then follows 
by comparing Eqs.~(\ref{outstat}) and (\ref{scatt}).
The result is expressed in terms of the Fourier transform 
of $V$, which is given by
\begin{equation}\label{tildev}
\tilde V({\bf q}) = \frac{4\pi iP}{ q R} \sin\left(\frac{q_x R}{2}\right),
\end{equation}
with $P=ZR$ and the momentum exchange ${\bf q}=k \hat e_r -{\bf k}$. 
We note that 
\begin{eqnarray}
q   &=& |{\bf q}|= 2k|\sin[(\theta-\phi_{{\bf k}})/2]|,\\
q_x &=& - \sigma' q\sin[(\theta+\phi_{\bf k})/2],  \nonumber
\end{eqnarray}
where $\sigma'=\pm$ is the sign of $\sin[(\theta-\phi_{\bf k})/2]$.
The scattering amplitude in Born approximation then reads
\begin{equation}\label{bornscatt}
f(\theta,\phi_{\bf k}) = - \sqrt{\frac{k}{8\pi v_F^2}}
\tilde V({\bf q)} \ b(\theta-\phi_{\bf k}; E_{{\bf k},\sigma}),
\end{equation}
where  
\begin{equation}\label{bdef}
b(\varphi;E) = \sum_\pm e^{\pm i\varphi/2} \left|\frac{E+\Delta}{E-\Delta}\right|^{\pm 1/2}
\end{equation}
is specific for Dirac fermions \cite{novikov}.
For $|E|\gg \Delta$, Eq.~(\ref{bdef}) simplifies to 
$b=2\cos(\varphi/2)$, reflecting the absence of backscattering \cite{rmp1}
for Dirac fermions in graphene,  $b(\pi)=0$.  

In the long-wavelength regime, $kR\ll 1$, Eq.~(\ref{tildev}) reduces to 
\begin{equation}
\tilde V({\bf q}) \simeq - 2\pi i \sigma' P 
 \sin \left(\frac{\theta+\phi_{\bf k}}{2}\right).
\end{equation}
Assuming also $|E_{{\bf k},\sigma}|\gg \Delta$, 
which effectively corresponds to the gapless case, $\Delta=0$, 
the Born approximation yields
\begin{equation}\label{scattamp}
\left|f(\theta,\phi_{\bf k})\right|^2 = \frac{2\pi k P^2}{v^2_F} 
\cos^2\left(\frac{\theta-\phi_{\bf k}}{2}\right) 
\sin^2\left(\frac{\theta+\phi_{\bf k}}{2}\right).
\end{equation}
In this expression, the $\cos^2(\cdots)$ factor comes from the ``Dirac factor''
$b$ in Eq.~(\ref{bdef}), while the $\sin^2(\cdots)$ factor reflects
the angular dependence due to the anisotropic electric dipole potential.
The transport cross-section, $\Lambda_{\rm tr}$, and the total 
cross-section, $\Lambda$, for the massless case 
then follow from standard definitions,
\begin{eqnarray}\nonumber
\Lambda_{\rm tr} &=& \int_0^{2\pi} d\theta  \
\left[ 1-\cos(\theta-\phi_{\bf k}) \right] \
\left |f(\theta,\phi_{\bf k})\right|^2 \\ 
\label{transportcr}
&=& \frac{\pi^2 k P^2 }{2v_F^2},\\ 
\nonumber \Lambda &=& \int d\theta \ |f(\theta,\phi_{\bf k})|^2 = 
\left(1+2\sin^2\phi_{\bf k}\right) \Lambda_{\rm tr}.
\end{eqnarray}
Remarkably, the transport cross-section is independent of the incidence angle 
$\phi_{\bf k}$.  This implies that
the dipole-induced angular dependence is precisely compensated
by the ``Dirac factor'' in Eq.~(\ref{scattamp}), which is responsible for the 
absence of backscattering.
The angle-independent result for $\Lambda_{\rm tr}$ implies that,
as long as the Born approximation is justified, the 
electrical conductivity remains isotropic even in the presence of arbitrarily  
oriented static electric dipoles.

\subsection{Nonperturbative analysis}\label{sec4d}

We next study the scattering problem for energies approaching
the band edges.   This requires a nonperturbative analysis, 
cf.~Ref.~\cite{garrett},
which we carry out in this subsection by adopting the point-like
dipole formulation in Eq.~(\ref{pointdipole}). For clarity,
we choose $\sigma=-1$, i.e., we write 
\begin{equation}\label{energdef}
E_{{\bf k}}=-\Delta+\epsilon, \quad \epsilon<0, \quad 
|\epsilon|\ll \Delta,
\end{equation}
in what follows.
The behavior near the other band edge, $\sigma=+1$, 
then follows by particle-hole symmetry. 
Far away from the nuclei, $r\gg R$, the two-center potential 
is well approximated by the point-like dipole, $V_d=-P\cos(\theta)/r^2$, 
and the Dirac equation reads
\begin{equation}\label{diracdip}
\left( \begin{array}{cc} V_d +2\Delta-\epsilon &
e^{-i\theta} (-i\partial_r - \frac{1}{r} \partial_\theta ) \\
 e^{i\theta} (-i\partial_r + \frac{1}{r} \partial_\theta ) &
V_d -\epsilon \end{array} \right) \left( \begin{array}{c}
\eta \\ \chi \end{array} \right)=0.
\end{equation}
To regularize the fall-to-the-center singularity for the $1/r^2$ potential,
we impose a boundary condition that forbids particle flow into a disk of radius
$r_0$ around the origin, with a short-distance scale $r_0\approx R$. 
In fact,  by comparing to the solution of the full two-center problem
\cite{dipolPRL}, one finds that the universal bound-state spectrum in 
Sec.~\ref{sec4a} is fully recovered from the point-like dipole
form with the choice $r_0= R/4$. 
 
Importantly, the radial and the angular parts can 
now be separated by mapping the Dirac equation to an equivalent 
2D Schr\"odinger equation. This is a controlled 
approximation for $P\ll \Delta r_0^2$ and energies near the band edge,
$|\epsilon|\ll \Delta$, where the upper spinor component is
always small compared to the lower one,
\begin{equation}\label{etacomp}
\eta(r,\theta) \simeq \frac{e^{-i\theta}}{2\Delta}
\left(i\partial_r + \frac{1}{r}\partial_\theta\right)\chi (r,\theta).
\end{equation}
Under these conditions, Eq.~(\ref{diracdip}) reduces to an 
effective 2D Schr\"odinger equation 
for the lower spinor component only,
\begin{equation}\label{schr}
\left[ -\frac{1}{2\Delta}\left(\partial_r^2+\frac{1}{r}\partial_r
+\frac{1}{r^2}\partial_\theta^2\right) +\frac{P\cos\theta}{r^2}+
\epsilon \right] \chi(r,\theta) = 0.
\end{equation}
The above-mentioned boundary condition at $r=r_0$ then 
implies a Dirichlet condition for the Schr\"odinger wavefunction,
i.e.,  $\chi=0$ for $r<r_0$. 

Fortunately, Eq.~(\ref{schr}) can now be separated
by the \textit{Ansatz} $\chi(r,\theta) = R(r) Y(\theta)$.
With the separation constant $\gamma$, the angular function
obeys a Mathieu equation, 
\begin{equation}\label{angular}
\left(\frac{d^2}{d\theta^2} + \gamma-2P\Delta\cos\theta \right)Y(\theta)=0,
\end{equation}
where $2\pi$-periodic solutions exist only when 
$\gamma$ matches one of the characteristic values \cite{gradst} 
of the Mathieu
equation, $\gamma=\gamma_{j,\kappa}(P\Delta)$.
Here, $\kappa=\pm$ is the parity of the solution, 
$Y_{j,\kappa}(-\theta)= \kappa Y_{j,\kappa}(\theta),$
and $j=0,1,2,\ldots$ effectively replaces the conventional
angular momentum, with $j+\kappa\ge 0$.  
The quantum numbers $(j,\kappa)$ have already appeared in
Sec.~\ref{sec4a}, where we discussed the bound-state spectrum,
with $\epsilon>0$ in Eq.~(\ref{energdef}).
Indeed, the angular equation (\ref{angular}) is independent of
the particle energy.  
Following standard notation \cite{gradst,abramowitz},
with the Mathieu functions ${\rm ce}_{2j}$ and ${\rm se}_{2j}$, and their
respective eigenvalues $a_{2j}$ and $b_{2j}$, the solutions to
Eq.~(\ref{angular}) are
\begin{eqnarray}\label{mathieusol}
Y_{j,+}(\theta) &=& {\rm ce}_{2j}\left(\frac{\theta}{2}, 4P\Delta
\right) ,\quad \gamma_{j,+}= \frac{1}{4}
a_{2j}(4P\Delta), \\  \nonumber
Y_{j,-}(\theta) &=& {\rm se}_{2j}\left(\frac{\theta}{2}, 4P\Delta
\right),\quad \gamma_{j,-}=\frac{1}{4} b_{2j}(4P\Delta).
\end{eqnarray}
For given dipole moment $P$, the characteristic values are ordered as
$\gamma_{0,+}<\gamma_{1,-}< \gamma_{1,+} <\gamma_{2,-}<\ldots,$
where $\gamma_{j,\kappa}=-s_{j,\kappa}^2<0$ for $P>P_{j,
\kappa}$, with $P_{j,\kappa}$ in Eq.~(\ref{pjk}) [note that
 $P_{0,+}=0$] and $s_{jk}$ in Eq.~(\ref{ssdef}).
For $P<P_{j,\kappa}$, on the other hand, the respective 
 Mathieu eigenvalue is positive, $\gamma_{j,\kappa}>0$.
With the solution of the angular equation at hand, 
the radial equation resulting from Eq.~(\ref{schr}) becomes a Bessel equation,
\begin{equation}\label{radial}
\left(\frac{d^2}{dr^2} + \frac{1}{r} \frac{d}{dr} - 
\frac{\gamma_{j,\kappa}}{r^2}+k^2
 \right) R_{{\bf k},j,\kappa}(r) = 0,
\end{equation}
where ${\bf k}$, with absolute value $k=\sqrt{-2\Delta \epsilon}$, denotes 
the incoming momentum of the scattering
state.  Note that up to this point, the above equations also
allow one to study bound-state solutions, see Sec.~\ref{sec4a} and 
Ref.~\cite{dipolPRL}.  The radial equation now contains a dependence
on the dipole moment only through the characteristic values of the Mathieu
equation.

The general solution of Eq.~(\ref{radial}) can be written in
terms of Hankel functions.  With complex coefficients 
$\beta_{{\bf k},j,\kappa}$, we obtain
\begin{equation}
R_{{\bf k}, j,\kappa}(r) \sim \beta_{{\bf k},j,\kappa} 
H_{\sqrt{\gamma_{j,\kappa}}}^{(1)} (kr) +H_{\sqrt{\gamma_{j,\kappa}}
}^{(2)} (kr),
\end{equation}
where $\sqrt{\gamma_{j,\kappa}}\to is_{j,\kappa}$ for $P>P_{j,\kappa}$,
see Eq.~(\ref{ssdef}).
For given quantum numbers $({\bf k},j,\kappa)$ characterizing the state, 
the Dirichlet condition at $r=r_0$ now fixes the $\beta$ coefficients.
From now on, we shall focus on the long wavelength regime, $kr_0\ll  1$,
where the short-distance form of the Hankel functions yields
\begin{equation}\label{betadef}
\beta_{{\bf k},j,\kappa}\simeq \left\{ \begin{array}{cc} 1 , & P<P_{j,\kappa}\\
e^{-\pi s_{j,\kappa}} \frac{ 
\sin \left[s_{j,\kappa}\ln (ikr_0/2) - \varphi(s_{j,\kappa})\right]
}{
\sin \left[ s_{j,\kappa}\ln (-ikr_0/2) - \varphi(s_{j,\kappa})\right]
},& P>P_{j,\kappa},\end{array}\right.
\end{equation}
with $\varphi(s)={\rm arg}\Gamma(1+is)$.
We have thereby constructed the nonperturbative scattering solution of the 
Dirac equation for $V=V_d$, which holds 
for energies near the (lower) band edge.  For the
lower spinor component, we find
\begin{equation}\label{gensol}
\chi(r,\theta) =  \sum_{j,\kappa} 
c_{j,\kappa} \left[ \beta_{{\bf k},j,\kappa} 
H_{\sqrt{\gamma_{j,\kappa}}}^{(1)} (kr) +H_{\sqrt{\gamma_{j,\kappa}}
}^{(2)} (kr)\right] Y_{j,\kappa}(\theta) ,
\end{equation}
with complex coefficients $c_{j,\kappa}$.
The upper spinor component, $\eta(r,\theta)$,  follows by virtue
of Eq.~(\ref{etacomp}).  The next step is to choose 
the $c_{j,\kappa}$ to match the asymptotic behavior 
of Eq.~(\ref{gensol}) to the general scattering state (\ref{scatt}),
which then determines the nonperturbative 
scattering amplitude $f(\theta,\phi_{\bf k})$.  

To that end, we first expand the incoming plane wave 
in terms of Mathieu functions. Employing the asymptotic
form of the radial solution, some algebra yields
\begin{eqnarray}
&& e^{ikr\cos(\theta-\phi_{\bf k})} \simeq 
\sqrt{\frac{2}{\pi i k r }} \sum_{j,\kappa} Y_{j,\kappa}(\theta) \\
\nonumber &&\times
\left[ Y_{j,\kappa}(\phi_{\bf k}) e^{ikr} +i Y_{j,\kappa}(\phi_{\bf k}+\pi)
e^{-ikr}\right].
\end{eqnarray} 
This implies that the coefficients $c_{j,\kappa}$ in the 
scattering state (\ref{gensol}) have to be chosen as
\begin{equation}
c_{j,\kappa} = e^{-i(\pi/2)\sqrt{\gamma_{j,\kappa}}} Y_{j,\kappa}(\phi_{\bf k}
+\pi).
\end{equation}
The scattering amplitude in Eq.~(\ref{scatt}) is therefore given by
\begin{eqnarray}\label{fjkfinal}
f(\theta, \phi_{\bf k}) &=& -i \sqrt{\frac{2}{\pi k}} \sum_{j,\kappa}
Y_{j,\kappa}(\theta) \\ \nonumber
&\times& \left[ \beta_{{\bf k},j,\kappa} 
e^{-i\pi\sqrt{\gamma_{j,\kappa}}} Y_{j,\kappa}(\phi_{\bf k}+\pi)
-Y_{j,\kappa}(\phi_{\bf k})\right].
\end{eqnarray}
For $P\Delta\to 0$, this result for the scattering ampltiude does not vanish, 
as may  have been expected since the dipole potential is then absent. 
However, our Dirichlet condition implies the (artificial) presence
of an infinitely repulsive hard-wall  potential at $r=r_0$, which 
produces a finite (but spurious) contribution to the scattering amplitude.
We have checked that for $P\Delta\to 0$, Eq.~(\ref{fjkfinal}) 
recovers the corresponding isotropic result for the impenetrable radial wall
potential, where the scattering amplitude depends
only on $\theta-\phi_{\bf k}$. We stress that there is a separate dependence
on $\theta$ and $\phi_{\bf k}$ in the presence of the dipole. 
However, for all scattering channels $(j,\kappa)$  not hosting
bound states, i.e., as long as $P<P_{j,\kappa}$  and therefore
$\gamma_{j,\kappa}>0$, the choice of the boundary
condition is immaterial and one can send $r_0\to 0$. 

\begin{figure}
\begin{center}
\vspace*{0.5cm}
\epsfxsize=0.96\columnwidth
\epsffile{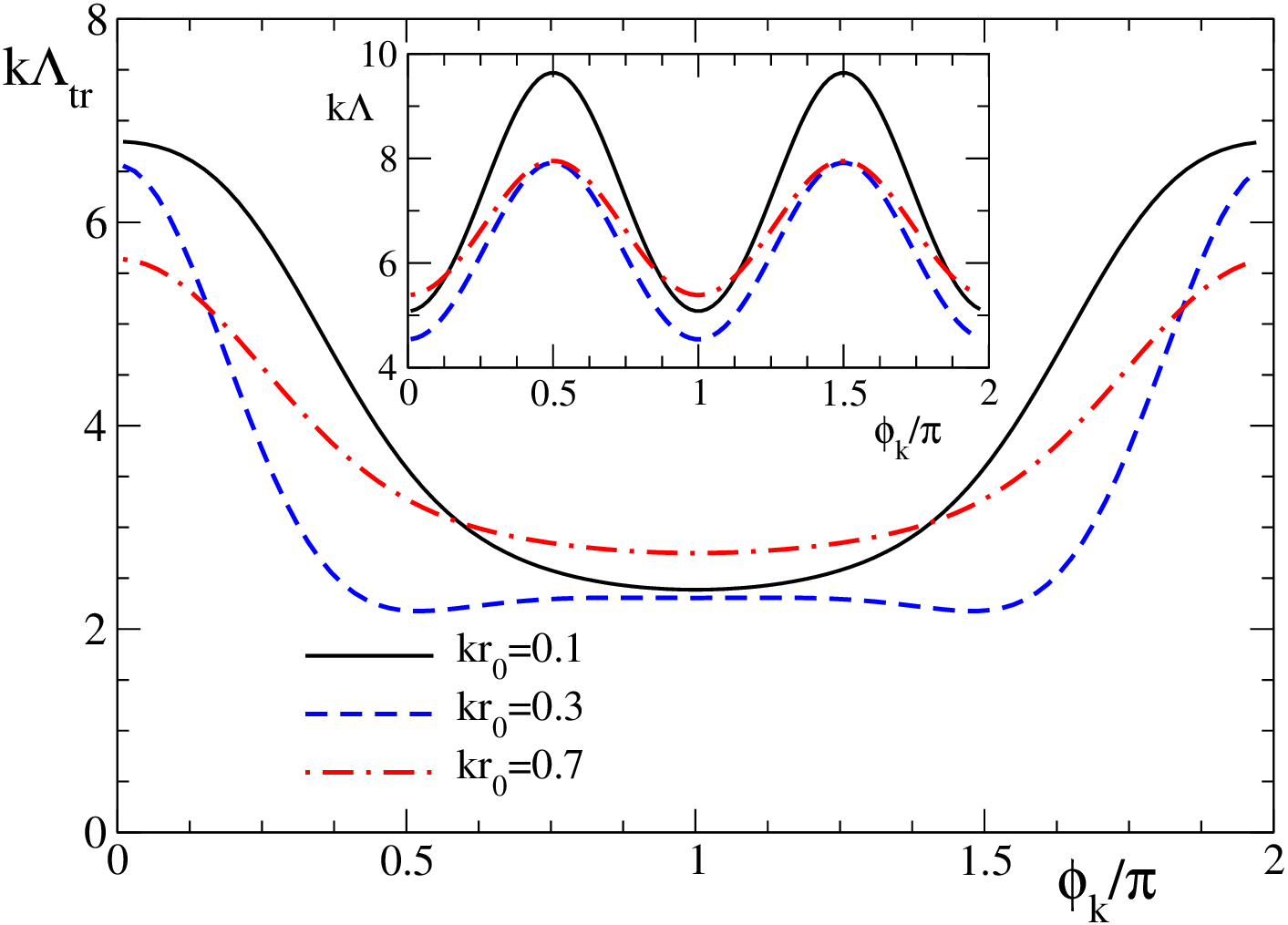}
\caption[]{Main panel: Angular dependence of the transport cross-section
$\Lambda_{\rm tr}$ vs incidence angle $\phi_{\bf k}$. The shown results are
for dipole moment $P\Delta=1.05$ and various values of $kr_0$.
They follow by numerical integration, see Eq.~(\ref{transportcr}), 
using the nonperturbative scattering amplitude (\ref{fjkfinal}). 
The lowest five $(j,\kappa)$ scattering channels have been included,
and $r_0$ can be identified with $R/4$, where $R$ is the distance
between the Coulomb centers.
Inset: Same but for the total cross-section $\Lambda$.
\label{fig3}}
\end{center}     
\end{figure}

The nonperturbative phenomena of main interest in this subsection 
involve scattering channels with dipole-induced bound states, i.e., 
 $P>P_{j,\kappa}$.  The scattering amplitude (\ref{fjkfinal}) then 
determines the transport cross-section, $\Lambda_{\rm tr}(\phi_{\bf k})$, 
and the total cross-section, $\Lambda(\phi_{\bf k})$,
according to the integrals in Eq.~(\ref{transportcr}).  Evaluating
these integrals numerically, we show typical results in Fig.~\ref{fig3},
where the lowest five scattering channels are included.  For the
shown results, the hard-core contribution due to the Dirichlet boundary
condition is negligible against the dipole-induced scattering.  
The total cross-section in the inset of Fig.~\ref{fig3} exhibits 
a very similar, $\pi$-periodic, angular
dependence as the Born approximation result in
Eq.~(\ref{transportcr}). 
However, in marked contrast to the prediction 
of the Born approximation, the nonperturbative result for the
transport cross-section clearly depends on the incidence angle $\phi_{\bf k}$. 
This effect can be traced back to the presence of dipole-induced bound states,
and directly implies that charge transport properties will be 
angle-dependent at energies approaching the edge, 
$|E_{{\bf k},\sigma}|\to \Delta$, where the Born approximation breaks down.  

\section{Conclusions}
\label{conc}

In this paper, we have discussed several noteworthy features of Dirac
fermions in graphene in the presence of a two-center potential.
For equal nuclear charges of slightly subcritical value, 
one can induce a transition to the supercritical regime by
lowering the distance $R$ between the Coulomb centers 
below a critical value $R_{\rm cr}$.   Our LCAO predictions
for the ground-state energy are qualitatively similar to previous results 
obtained by an asymptotic matching approach \cite{gusynin}.   

For opposite charges, the potential at large distances is 
equivalent to a static electric dipole potential.  
In graphene, even a very weak dipole can 
capture infinitely many bound states, and we have 
addressed the corresponding scattering problem in some detail.
For energies not too close to the band edge, the Born 
approximation is valid and predicts that the transport 
cross-section is isotropic.  This conclusion can be rationalized
by noting that the dipolar angular dependence is precisely compensated
by the one due to the Dirac nature of the quasi-particles in graphene. 
Important deviations from the Born approximation originate
from scattering channels that are linked to bound states.
Note that there is at least one infinite tower of bound states for 
arbitrary dipole strength.  We have determined a nonperturbative
solution for the scattering amplitude within a point-like dipole
model, which indicates that a nontrivial angular dependence 
of the transport cross-section will be present as a consequence of  
such effects.

To conclude, we hope that our predictions can soon be probed
experimentally by scanning tunneling spectroscopy on graphene
monolayers along the lines of Refs.~\cite{exp1,exp2,exp3}.

\vspace{1cm}

We thank A.~Altland, E.~Andrei, J.-C.~Cuenin, H.~Siedentop,
and A.~Zazunov for valuable discussions. Financial support by the 
DFG (SFB TR12 and SPP 1459) and by the Volkswagen-Stiftung is 
gratefully acknowledged.

\end{document}